\documentclass[3p,preprint,12pt]{elsarticle}
\usepackage{graphics}
\usepackage{ecrc}
\usepackage{amssymb}

\usepackage{adjustbox}

\volume{00}

\firstpage{1}

\journalname{Nuclear Physics A}

\runauth{}

\jid{procs}

\jnltitlelogo{Nuclear Physics A}

\journal{Nuclear Physics A}

\newcommand{\Cd}{$^{106}$Cd}
\newcommand{\Sn}{$^{110}$Sn}
\newcommand{\Snn}{$^{109}$Sn}
\newcommand{\In}{$^{109}$In}
\newcommand{\Carbon}{$^{12}$C}
\newcommand{\raa}{($\alpha,\alpha$)}

\newcommand{\rag}{($\alpha,\gamma$)}
\newcommand{\ran}{($\alpha,n$)}
\newcommand{\rap}{($\alpha,p$)}

\newcommand{\al}{$\alpha$}

\newcommand{\anucpots}{$\alpha$--nucleus potentials}

\begin{document}

\begin{frontmatter}

\title{The \Cd\raa\Cd\ elastic scattering in a wide energy range for $\gamma$ process studies}

\author[ATOMKI]{A. Ornelas}
\author[ATOMKI]{G. G. Kiss \fnref{fn1}\corref{cor1}}
\ead{ggkiss@atomki.mta.hu}
\cortext[cor1]{corresponding author}
\fntext[fn1]{Present Address: RIKEN Nishina Center, 2-1 Hirosawa, Wako, Saitama 351-0198, Japan }
\author[ATOMKI]{P. Mohr}
\author[Centro,FCUL]{D. Galaviz}
\author[ATOMKI]{Zs. F\"ul\"op}
\author[ATOMKI]{Gy. Gy\"urky}
\author[ATOMKI]{Z. M\'{a}t\'{e}}
\author[Hertfordshire,Basel]{T. Rauscher}
\author[ATOMKI]{E. Somorjai}
\author[goethe]{K. Sonnabend}
\author[cologne]{A. Zilges}

\address[ATOMKI]{Institute for Nuclear Research (MTA Atomki), H-4001 Debrecen, POB.51., Hungary}
\address[Centro]{Centro de F\'{i}sica Nuclear, University of Lisbon, 1649-003 Lisbon, Portugal}
\address[FCUL]{Faculdade de Ci\^encias da Universidade de Lisboa, FCUL, Edif\'{i}cio C8, Campo Grande, 1749-016 Lisboa, Portugal}
\address[Hertfordshire]{Centre for Astrophysics Research, School of Physics, Astronomy and Mathematics, University of Hertfordshire, Hatfield AL10 9AB, United Kingdom}
\address[Basel]{Department of Physics, University of Basel, CH-4056 Basel, Switzerland}
\address[goethe]{Institut f\"{u}r Angewandte Physik, Goethe-University Frankfurt, D-60438 Frankfurt, Germany}
\address[cologne]{Institut f\"{u}r Kernphysik, Universit\"{a}t zu K\"{o}ln, D-50937, K\"{o}ln, Germany}

\begin{abstract}
Alpha elastic scattering angular distributions of the \Cd \raa \Cd \ reaction were measured at three energies around the Coulomb barrier to provide a sensitive test for the $\alpha$ + nucleus optical potential parameter sets. Furthermore, the new high precision angular distributions, together with the data available from the literature were used to study the energy dependence of the locally optimized $\alpha$ + nucleus optical potential in a wide energy region ranging from E$_{Lab}$  = 27.0 MeV down to 16.1 MeV. 
 
The potentials under study are a basic prerequisite for the prediction of $\alpha$-induced reaction cross sections and thus, for the calculation of stellar reaction rates used for the astrophysical $\gamma$ process. Therefore, statistical model predictions using as input the optical potentials discussed in the present work are compared to the available $^{106}$Cd + alpha cross section data. 

\end{abstract}


\end{frontmatter}

\section{Introduction}
\label{sec:intro}

The bulk of the naturally occurring nuclei heavier than iron observed in the Solar System were synthesized via neutron capture processes. In the case of the so-called astrophysical \textsl{s} process, the neutron flux is moderate, elements are synthesized through slow neutron captures and $\beta$ decays up to $^{209}$Bi \cite{kap11, rei14}. On the contrary, in an astrophysical environment where the neutron densities are orders of magnitude higher than the ones available for the \textsl{s} process, the neutron captures become faster than the $\beta$-decays and extremely neutron-rich nuclides close to the neutron drip line can be synthesized. These nuclei decay back into the valley of stability when the neutron flux ceases. This process --- the so-called astrophysical \textsl{r} process --- can only take place in explosive environments \cite{Arno14, Arg04}.

Furthermore, there are about 30-35 proton-rich nuclides between $^{74}$Se and $^{196}$Hg, which cannot be formed by neutron capture processes \cite{Woos78,Arno03, Kus11}. They are the so-called \textsl{p} isotopes. The number  of \textsl{p} isotopes depends on the one hand on the state of art \textsl{s} process models: a recent investigation \cite {arl99} showed that there are large \textit{s} process contributions to the abundance of $^{164}$Er, $^{152}$Gd and $^{180}$Ta. On the other hand, the precise estimation of the \textsl{r} process contribution to the abundance of certain \textsl{p} isotopes (e.g. $^{113}$In or $^{115}$Sn, see in \cite{nem94}) is essential, too. Consequently the modification of the list of \textsl{p} isotopes could be necessary.

The natural isotopic abundance of the \textsl{p} isotopes is typically 10 to 100 times less than that of the more neutron-rich isotopes of the same element that were created in the \textsl{s} or \textsl{r} processes. In their production, photon-induced reactions at temperatures around a few GK play a crucial role. It is generally accepted that the main stellar mechanism synthesizing these nuclei --- the so-called $\gamma$ process \cite{Rau13a} --- is initiated by photodisintegration reactions on pre-existing neutron-rich \textsl{s} and \textsl{r} seed nuclei. Photons with high energy and high flux, necessary for the photodisintegrations, are available in explosive nucleosynthesis scenarios, as in the Ne/O burning layer in Type II- \cite{Woos78,Arno03,Kus11} or in the Type Ia supernovae \cite{Trav11,Trav14}. 
However, based solely on this model, the calculations are not able to reproduce the Solar System \textsl{p} abundances mainly in two mass regions: 

--- The light \textsl{p} isotopes near the Mo-Ru region are largely underproduced by the photodisintegration reactions. In recent years, several other processes have been discussed that could contribute to the production of light \textsl{p} nuclei through proton captures in a highly proton-rich material: the so-called \textsl{rp}-(rapid proton capture) process \cite{Scha98}, $\nu$p-process \cite{Froh06a}, and proton captures on highly enriched \textsl{s} process seeds during Type Ia supernova explosion \cite{Trav11, Trav14, Howa91}. However, the role of these processes in nucleosynthesis remains unclear, e.g., for the \textsl{rp}-process it is uncertain whether the produced nuclei can be expelled into the interstellar medium and plays any role in the galactic chemical evolution. 

--- The \textsl{p} nuclei in the intermediate region between 150 $\leq$ A $\leq$ 165 are underproduced, too \cite{Raus02,Raye95}.  It remains unclear whether this deficiency is due to nuclear cross sections, stellar physics, or if alternative / additional process has to be invoked.

Although the basics of the $\gamma$ process were laid down several decades ago, many details of the process are still unknown.  
On the one hand, this concerns the ambiguities in the astrophysical conditions under which the process takes place (seed isotope abundances, peak temperatures, time scale, etc). On the other hand, large uncertainties are introduced into the calculations by the nuclear physics input, most importantly by the reaction rates (determined from cross sections). All $\gamma$ process models require the use of huge reaction networks including tens of thousands of nuclear reactions, and the rates of these reactions at a given stellar temperature are necessary inputs to the network calculations. The reaction rates are generally taken from calculations using the Hauser-Feshbach (H-F) statistical model \cite{Raus97,Raus00}. The accuracy of the H-F predictions mainly depends on the adopted nuclear models for optical-model potentials, $\gamma$-ray strength functions, and nuclear-level densities. Previous $\gamma$ process related experiments showed that above the A $\approx$ 100 region the precise knowledge on the $\alpha$ + nucleus optical potential would be highly desirable \cite{Kiss11b,Raus12,Saue11,Nett13,Kiss14,Glo14}, in order to increase the predictive power of the model calculations.

The subject of the present work is the study of the angular distributions of elastically scattered alpha particles on the $^{106}$Cd \textsl{p} nucleus. On the one hand experimental angular distributions measured at higher energies are available in the literature \cite{Palu12} and therefore, the opportunity to study the energy dependence of the imaginary part of the $\alpha$ + nucleus optical potential in a wide energy region ranging from E$_{Lab}$ = 27 MeV down to E$_{Lab}$ = 16.1 MeV is given. Furthermore, recently the cross sections of the $\alpha$-induced reactions on $^{106}$Cd were measured at very low energies, close to the astrophysically relevant ones \cite{Gyur06}. It was found that there is about a factor of 3 difference between the experimental cross sections and the H-F predictions in the case of the \rag\ reaction. The inaccurate knowledge on the $\alpha$ + nucleus optical potential was identified as the source of this disagreement.

Finally, the elastic scattering data obtained extends the already existing data set of experiments performed on several proton-rich nuclei \cite{Palu12,Mohr13,Mohr97,Fulo01,Gala05,Kiss06,Kiss08,Kiss09,Kiss11,Kiss13} and provides additional data to test the global parametrizations used in astrophysical network calculations.

The paper is organized as follows. Experimental details are described in Sec.~\ref{sec:exp_proc}. The measured angular distributions are presented in Sec.~\ref{sec:res}. Here, the new data are analyzed using a locally adjusted potential, and predictions from global \anucpots\ are shown. In Sec.~\ref{sec:capture_reac}, we compare the results of the H-F calculations --- calculated with the use of the local and global potentials discussed in the present work --- with the available $\alpha$-induced cross section data taken from the literature \cite{Gyur06}. Finally, conclusions are drawn in Sec.~\ref{sec:summary}. 

\section{Experimental procedure}
\label{sec:exp_proc}

The experiments were performed at the cyclotron laboratory of Atomki, in Debrecen, Hungary. The alpha elastic scattering angular distributions between $\vartheta_{Lab}$ = 20$^{\circ}$ and $\vartheta_{Lab}$ = 170$^{\circ}$ were measured on the \textsl{p} nucleus \Cd\ at three energies around the Coulomb barrier. Highly enriched $^{106}$Cd targets ($\sim 97\%$) of 250 $\mu$g/cm$^2$ thickness were prepared using vacuum evaporation technique. The experimental setup was similar to that used in previous experiments~\cite{Kiss13} (and references therein). In the following paragraphs, a short outline of the most relevant characteristics of the experimental setup is given. A detailed description of the experimental setup can be found in e.g. \cite{Gala04}.

The angular distributions of the elastically scattered alpha particles were measured with the help of two pairs of silicon detectors, mounted on two independently rotatable turntables. The solid angles of the detectors were between $1.63 \times 10^{-4}$ sr and $1.55 \times 10^{-4}$ sr. In addition, two silicon detectors under $\vartheta_{Lab}$ = 15$^\circ$, with a solid angle of $8.10 \times 10^{-6}$ sr were used for absolute normalization, to monitor the target thickness and to take into account the beam position on the target. 

\begin{figure}[ht!]
\begin{center}
\includegraphics[ width =0.95\textwidth , clip]{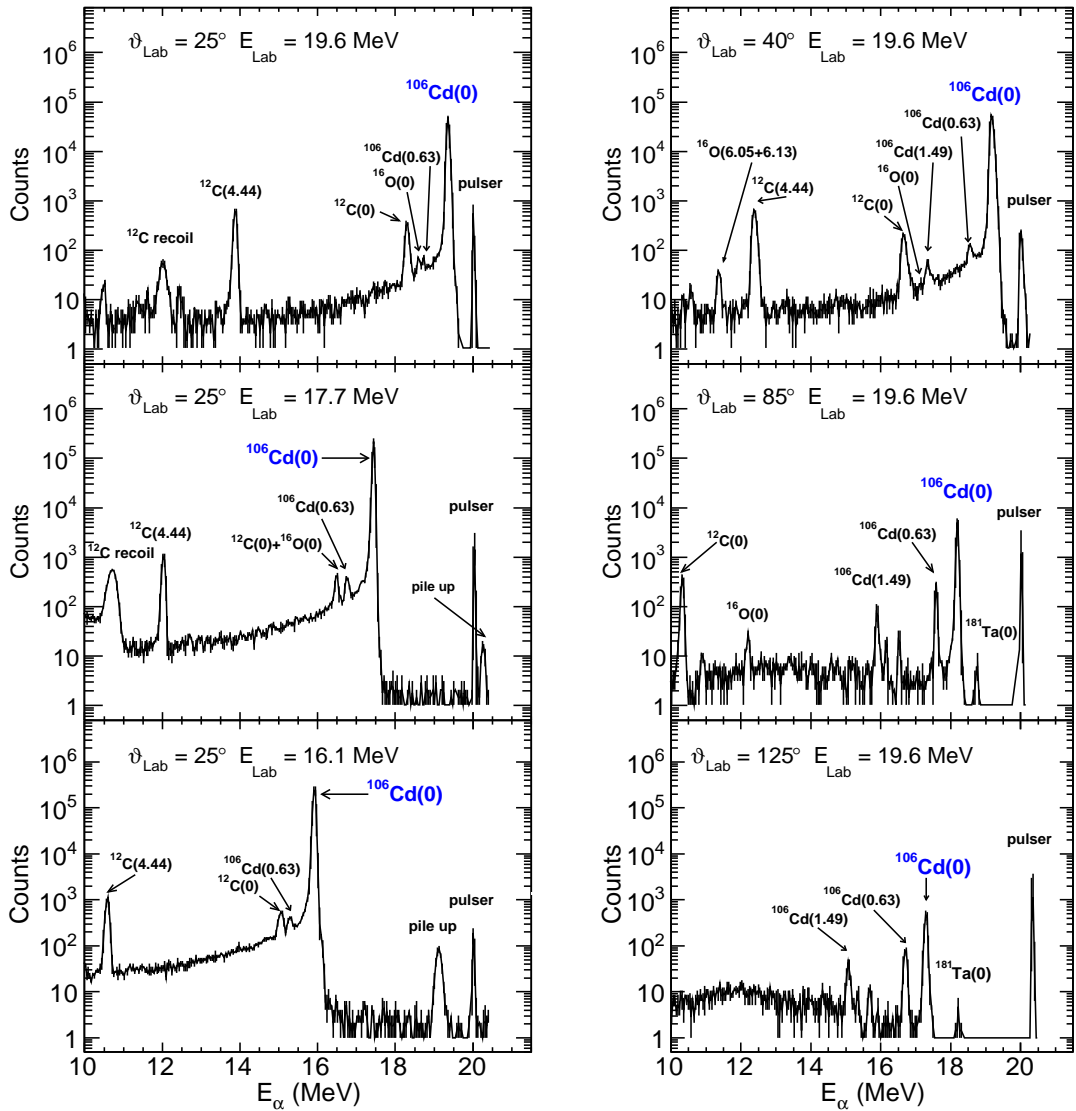}
\caption{
\label{fig:spectra}
(Colour online) Typical spectra taken for the reaction \Cd \raa \Cd . On the left panels, the spectra taken at a fixed angle $\vartheta_{Lab}$ = 25$^\circ$ are displayed for all three measured energies. The right side of the figure shows spectra at forward, middle, and backward angles for a fixed energy E$_{Lab}$ = 19.6 MeV. The elastic peak can be clearly distinguished from other peaks in all cases. In each spectrum, the pulser signals used to control the dead time were set at energies higher than the elastic peak. 
The most prominent peaks are labeled with the corresponding excitation energies of the residual nuclei. 
}
\end{center}
\end{figure}

The energies of the alpha particle beam were E$_{Lab}$ = 16.1 MeV, 17.7 MeV and 19.6 MeV with typical beam intensities in the order of 250 nA.
Two different sets of spectra registered during the measurement of the elastic scattering cross section are shown in Fig~\ref{fig:spectra}. On the left side of the figure, the spectra measured at position $\vartheta_{Lab}$ = 25$^\circ$ are shown for the three measured energies. The right side of the figure presents the spectra recorded at E$_{Lab}$ = 19.6 MeV, for three different angular positions. As it can be seen, the elastic peak can be easily separated from the additional components, both at forward (carbon and oxygen peaks) and backward (inelastic scattering on \Cd) angles. In addition, pulser signals were fed into the pre-amplifiers in order to have a control on the dead-time of the detection system. 
The pulser signal amplitude was set so that the pulser peak always falls above the \Cd\ elastic peak (see Fig. \ref{fig:spectra}).

\begin{figure}[ht!]
\begin{center}
\includegraphics[width =0.45\textwidth, clip]{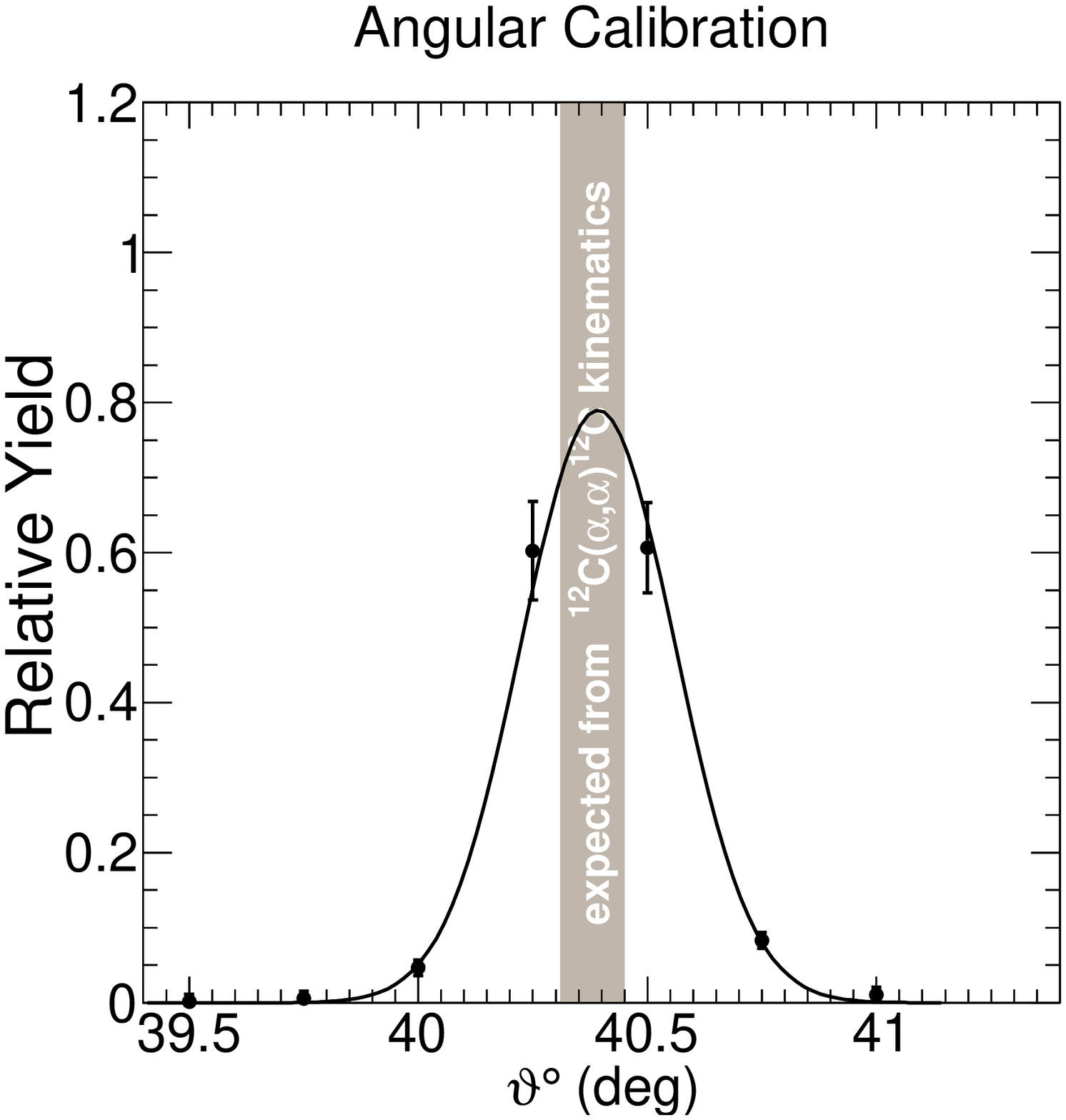}
\caption{
\label{fig:ang_calib}
Results of the coincidence measurement. The relative number of \Carbon\ recoils measured in coincidence with the elastically scattered alpha particles ($\vartheta_{\alpha}$ = 80$^\circ$) is shown as a function of the recoil angle. The gray region shows the expected position for the maximum rate. The solid line is a Gaussian curve to guide the eye. The accuracy of the measured angle was determined to be $\Delta \vartheta$ = 0.07$^\circ$.
}
\end{center}
\end{figure}

Knowledge of the exact angular position of the detectors is of crucial importance for the precision of a scattering experiment because the Rutherford cross section depends sensitively on the angle. The uncertainty in the Rutherford normalized cross section is dominated by the error of the scattering angles in the forward region. In a similar way to previous experiments~\cite{Kiss13} (and references therein) we took advantage of the kinematics of the reaction \Carbon \raa \Carbon\, and measured the elastically scattered alpha particles in coincidence with the recoil \Carbon. We fixed one of the detectors on one side of the chamber at $\vartheta_{\alpha}$ = 80$^\circ$ to measure the alpha particles, and measured in coincidence the recoil \Carbon\, nuclei around the expected angle of $\vartheta_{^{12}C}$ = 40.38$^\circ$. The result can be seen in Fig.~\ref{fig:ang_calib}. The uncertainty of the scattering angle was determined to be $\Delta \vartheta$ = 0.07$^\circ$. Several measurements were performed at other angular positions, and confirmed this result, ensuring the high precision of the experimental data. 

\section{Results and discussion}
\label{sec:res}
Complete angular distributions between 20$^{\circ}$ and 170$^{\circ}$ were measured at energies of E$_{Lab}$ = 16.1 MeV, 17.7 MeV and 19.6 MeV in 1$^{\circ}$ (20$^{\circ}$ $\leq$ $\vartheta$ $\leq$ 100$^{\circ}$), 1.5$^{\circ}$ (100$^{\circ}$ $\leq$ $\vartheta$ $\leq$ 140$^{\circ}$), and 2$^{\circ}$ (140$^{\circ}$ $\leq$ $\vartheta$ $\leq$ 170$^{\circ}$) steps.
The statistical uncertainties varied between 0.3\% (forward angles) and 1 to 2\% (backward angles) \cite{Kiss06}. The count rates $N(\vartheta)$ have been normalized to the yield of the monitor detectors $N_{\rm Mon.}$($\vartheta$ = 15$^{\circ}$):

\begin{eqnarray}
\left(\frac{d\sigma}{d\Omega}\right)(\vartheta) = \left(\frac{d\sigma}{d\Omega}\right)_{\rm Mon} \frac{N(\vartheta)}{N_{\rm Mon}} \frac{\Delta\Omega_{\rm Mon}}{\Delta\Omega}
\end{eqnarray}

The solid angle of the detectors is represented by $\Delta\Omega$. The Rutherford-normalized cross sections  cover around 2 orders of magnitude, taking into account all the angular distributions measured ($E_{Lab} = 16.1$ MeV to $E_{Lab} = 27.0$ MeV). Over this range of cross sections, our measured angular distributions have an accuracy of around 4 to 5\%, while the data in \cite{Palu12} claim around 1 to 2\% at forward angles and the authors add $ \leq 10$\% for backward angles due to the relative low enrichment. 

In the following, the analysis of the new experimental data in the framework of the optical model will be presented. Our analysis is extended up to E$_{Lab}$ = 27 MeV by taking into account the elastic scattering angular distributions from \cite{Palu12}. 

The optical model potential combines a Coulomb term with the complex form of the nuclear potential, which consists of a real and an imaginary part. Usually, the parameters of the optical potential are derived from the analysis of the angular distributions of elastically scattered alpha particles and are adjusted to experimental $\alpha$-induced cross sections if they are known. The $\alpha$ + nucleus optical potential is given by:

\begin{eqnarray}
U(r) = V_{\rm{C}}(r) + V(r) + iW(r)
\label{eq:pot}
\end{eqnarray}

where $V_{\rm{C}}(r)$ is the Coulomb potential (calculated from a homogeneously charged sphere), $V(r)$ and $W(r)$ are the real and imaginary parts of the nuclear potential, respectively. In the following, we will first determine a locally adjusted potential (Sec.~\ref{subsec:opt_mod}) and then compare the new experimental data to predictions of global $\alpha$ + nucleus optical potentials (Sec.~\ref{subsec:ang_theo}). 

\begin{figure}[ht!]
\begin{center}
\includegraphics[width =1.0\textwidth, height = 190mm, clip]{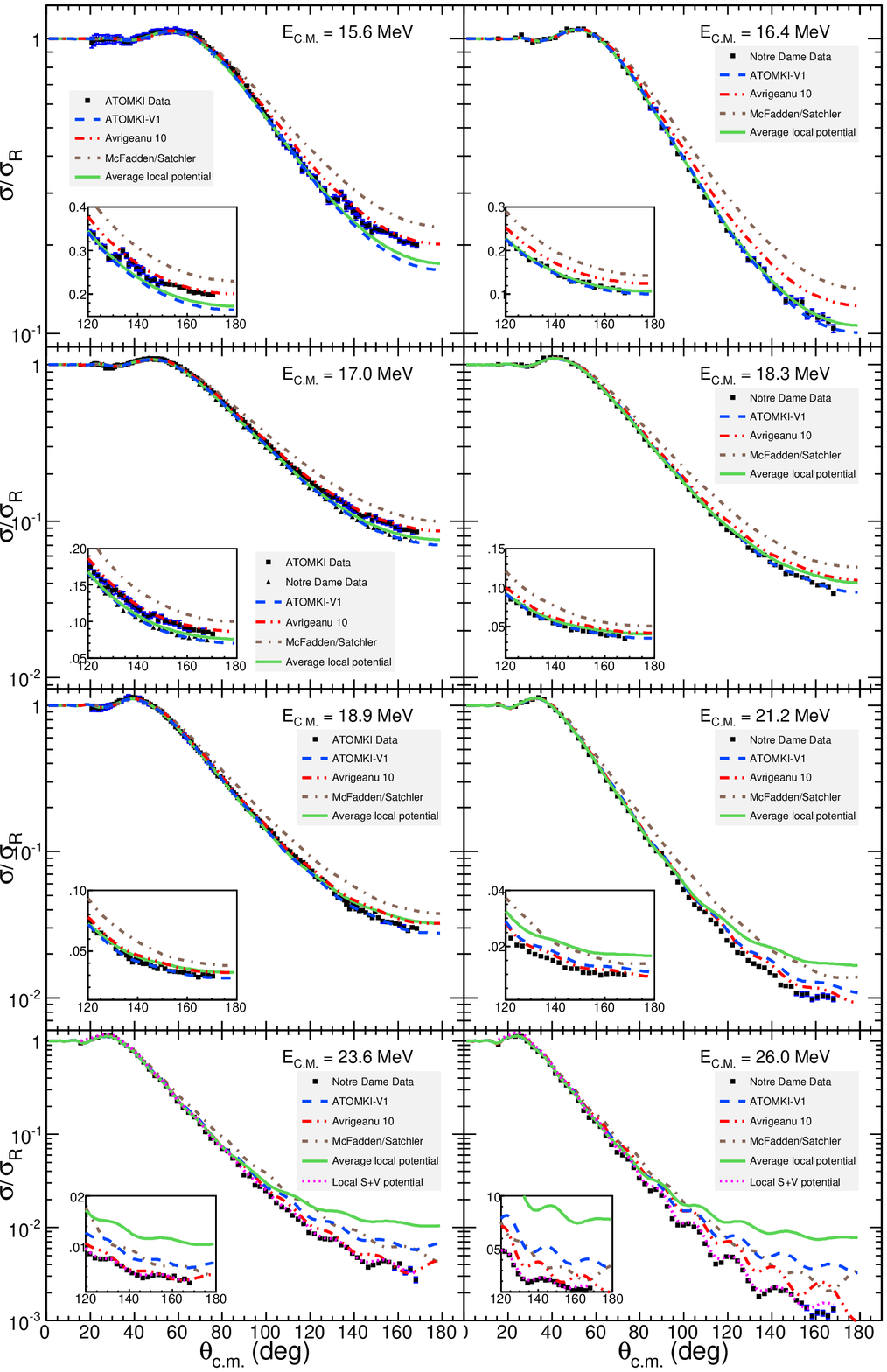}
\caption{
\label{fig:cd106_all_pot_ND}
(Colour online) Available experimental \Cd \raa \Cd\ elastic scattering cross sections --- measured at Atomki and at University of Notre Dame \cite{Palu12}---, normalized to the Rutherford cross section compared to theoretical predictions calculated using global $\alpha$ + nucleus optical potentials. For details, see text.} 
\end{center}
\end{figure}

\subsection{Local optical potential analysis}
\label{subsec:opt_mod}

A local optical potential is tailored to describe the experimental data of a small region of the mass range, i.e., it is a potential with parameters derived from the data of a few alpha elastic angular distributions or even from just a single experiment. This type of potential is used as an analysis tool, since it is possible to achieve $\chi^2_{red}$ very close to unity, and therefore it is possible to study the impact/sensitivity of its different parameters in the description of the experimental data.

\begin{table}[ht!]
\caption{
\label{tab:local_pot_parameters}
Parameters of the local optical potentials derived from the fits to the elastic scattering angular distributions measured at Atomki (Atomki) and at University of Notre Dame \cite{Palu12} (ND), for further details, see text.
}
\begin{adjustbox}{max width=\textwidth}
\begin{tabular}{llllllllllll}
\hline
{\bf $E_{C.M}$} & {\bf $\lambda$} & {\bf $w$} & {\bf $J_R$} & {\bf $r_{rms,R}$} & {\bf $W_S$} & {\bf $R_S$} & {\bf $a_S$} & {\bf $J_I$} & {\bf $r_{rms,I}$} & {\bf $\sigma_{reac}$} & {\bf $\chi^2_{red}$} \\
(MeV) & (-) & (-) & (MeV fm$^3$) & (fm) &(MeV) & (fm) & (fm) & (MeV fm$^3$) & (fm) & (mb) & (-) \\
\hline
15.6 - Atomki & 1.218 & 1.029 & 351.84 & 5.3602 & 193.01 &	1.563 & 0.291 & 91.66  & 7.490  & 341.9 & 0.8 \\
16.4 - ND  & 1.211 & 1.020 & 341.06 & 5.3152 & 169.47 & 1.468 & 0.413 & 101.19   & 7.1421 & 491.3 & 1.0 \\ 
17.0 - Atomki & 1.373 & 1.004 & 368.91 & 5.2324 & 166.30 & 1.464 & 0.403 & 96.51  & 7.1158 & 544.4 & 0.9 \\ 
17.0 - ND  & 1.423 & 1.007 & 385.38 & 5.2462 & 171.17 & 1.446 & 0.436 & 105.02   & 7.0679 & 581.2 & 0.7 \\ 
18.3 - ND  & 1.371 & 1.005 & 369.37 & 5.2370 & 149.35 & 1.428 & 0.464 & 95.30  & 7.0147 & 733.4 & 1.1 \\
18.9 - Atomki & 1.378 & 1.000 & 365.29 & 5.2076 & 138.73 & 1.402 & 0.505 & 93.26  & 6.9422 & 803.0 & 0.8 \\ 
21.2 - ND  & 1.380 & 1.008 & 374.43 & 5.2500 & 134.37 & 1.419 & 0.500 & 91.41  & 7.0108 & 1039.5  & 2.0 \\ 
21.2 - ND$^{*}$& 1.393 & 1.004 & 373.74 & 5.2304 & 137.40 & 1.414 & 0.492 & 91.26  & 6.9812 & 1020.9 & 1.7 \\ 
\hline
Average    & 1.336 & 1.010 & 365.18 & 5.2641 & 160.34 & 1.456 & 0.430 & 96.34  & 7.1119 &  -  & - \\
Av. w/ Ren.& 1.338 & 1.010 & 365.08 & 5.2613 & 160.78 & 1.455 & 0.429 &	96.31  & 7.1077 &  -  & - \\
\hline
23.6 - ND  & 1.401 & 0.994 & 364.39 & 5.1760 & 159.16 & 1.372 & 0.492 & 99.72  & 6.7903 & 1150.6  & 5.4 \\ 
23.6 - ND$^{*}$& 1.354 & 1.007 & 366.90 & 5.2475 & 152.59 & 1.388 & 0.513 & 102.18   & 6.8891 & 1209.2  & 2.6 \\ 
26.0 - ND  & 1.446 & 0.989 & 370.83 & 5.1515 & 132.06 & 1.372 & 0.512 & 86.21  & 6.8157 & 1287.2  & 13.2  \\ 
26.0 - ND$^{*}$& 1.392 & 1.004 & 373.17 & 5.2288 & 128.98 & 1.392 & 0.525 & 88.96  & 6.9209 & 1346.5  & 7.4 \\ 
\hline
\multicolumn{12}{l}{$^{*}$ - Abs. renormalization: 1.016 at 21.2 MeV ; 0.946 at 23.6 MeV ; 0.929 at 26.0 MeV} \\
\hline
23.6 - S+V $\dag$ & 1.433 & 0.990 & 368.14 & 5.1547 &7.65& 1.686 & 0.423 & 54.10  & 4.8021 & 1153.2 & 1.4 \\  
\multicolumn{4}{c}{W$_V$ = 31.45} & \multicolumn{4}{c}{R$_V$ = 1.131} & \multicolumn{4}{c}{a$_V$ = 0.145}  \\ 
26.0 - S+V $\dag$ & 1.413 & 0.996 & 370.10 & 5.1880 &10.65 & 1.648 & 0.464 & 54.82  & 5.0561 & 1309.1 & 2.5 \\ 
\multicolumn{4}{c}{W$_V$ = 29.98} & \multicolumn{4}{c}{R$_V$ = 1.144} & \multicolumn{4}{c}{a$_V$ = 0.162}  \\ 
\hline
\multicolumn{2}{l}{21.2 - Phase Shift } & & & &       &       &       &          &        & 1006.4 & 1.2 \\ 
\multicolumn{2}{l}{23.6 - Phase Shift $\dag$} & & & &       &       &       &          &        & 1154.7 & 1.1 \\
\multicolumn{2}{l}{26.0 - Phase Shift $\dag$} & & & &       &       &       &          &        & 1309.1 & 1.2 \\
\hline
\multicolumn{12}{l}{$\dag$ - Abs. renormalization: 0.959 at 23.6 MeV ; 0.921 at 26.0 MeV} \\
\hline
\\
\end{tabular}
\end{adjustbox}
\end{table}

The real part of the local potential is based on the double folding procedure of \cite{Abel93} calculated using charge densities based on electron scattering \cite{Vrie87}; this $V(r)$ potential is modified by a strength parameter $\lambda$ and a width parameter $w$, as in the case of the global potential ATOMKI-V1 (see Sec. \ref{subsec:ang_theo}). The nucleon-nucleon interaction is parametrized using the well known DDM3Y interaction \cite{Mohr97,Satc79,Atzr96}. The fitted local potential has a Woods-Saxon form imaginary part $W(r)$, which consists of a surface $W_{S}$ term only at very low energies \cite{Avri09}.

By fitting the data from each of the angular distributions, the $W_S$, $R_S$ and $a_S$ parameters of the imaginary part were left as free parameters for the fitting. For the real part the bare double folding potential was calculated by the DFOLD code \cite{DFOLD}, using the aforementioned DDM3Y interaction, and was fixed for all the $E_{Lab}$. The width $w$ and strength $\lambda$ parameters of the real part were left free, in order to modify the bare folding potential (with $\lambda = w = 1$). 

The results of this local potential study are shown in Table \ref{tab:local_pot_parameters}. The table lists all the resulting parameters, as well as the calculations for the volume integrals $J_R$ and $J_I$, root mean-square radii $r_{rms,R}$ and $r_{rms,I}$, the reaction total cross section $\sigma_{reac}$ and the obtained $\chi^2_{red}$.

It is obvious from Table \ref{tab:local_pot_parameters} that the angular distributions above 20\,MeV measured at Notre Dame \cite{Palu12} can only be described with relatively poor $\chi^2_{\rm{red}}$. There are two reasons for this unexpected large $\chi^2/F$. 

First, the absolute values of the Notre Dame data \cite{Palu12} have been normalized to Rutherford scattering ($\sigma/\sigma_R = 1.0$) at the most forward experimental angle. However, this procedure is not justified at the higher energies. We have allowed to vary the absolute normalization of these angular distributions, leading to $\chi^2_{\rm{red}}$ which are about a factor of two smaller than the ones using the original normalization. The normalization factors are 1.016, 0.946, and 0.929 at 21.2\,MeV, 23.6\,MeV, and 26.0\,MeV, respectively.

Second, as already mentioned above, a surface imaginary potential is not sufficient at higher energies, and consequently a volume Woods-Saxon potential has been added (labelled ``S+V'' in Table \ref{tab:local_pot_parameters}). This improves the obtained $\chi^2_{\rm{red}}$ by another factor of two. 

As a further test, in addition to the above mentioned local potential fits, phase shift fits were also performed for the Notre Dame data \cite{Palu12} using the method of \cite{Chis96}. Here again $\chi^2_{\rm{red}} \approx 1$ can only be obtained with modified absolute normalization of the data, and the best-fit values for the absolute normalization agree within about 1\,\% with the local potential fits. This confirms that the absolute normalization of these data has to be revised.

For the prediction of low-energy $\alpha$-induced cross sections we have chosen an average potential from all angular distributions below 21.2\,MeV which can be described using a pure surface Woods-Saxon imaginary
potential. These average values are practically insensitive to the modified normalization of the Notre Dame data \cite{Palu12} because the strongly affected data at 23.6 MeV and 26.0\,MeV do not contribute here (the additional volume Woods-Saxon imaginary potential complicates the extrapolation to low energies). The minor renormalization of the 21.2\,MeV data affects the average local potential parameters by the order of 0.1\,\%. Unfortunately, our analysis is not sufficiently sensitive to determine the energy dependence of the parameters of the imaginary part.

Figure \ref{fig:cd106_all_pot_ND} presents the predicted angular distributions calculated using this average local potential for each of the recently measured elastic scattering cross sections, together with the experimental data taken at University of Notre Dame \cite{Palu12}, to illustrate the overall good description. For completeness, Fig. \ref{fig:cd106_all_pot_ND} shows in the case of E$_{c.m.}$ = 23.6 MeV and 26.0 MeV the local potential with additional volume contribution. The local fits using the parameters from Table \ref{tab:local_pot_parameters} and the phase shift fits are not shown in Fig.~\ref{fig:cd106_all_pot_ND} because they follow exactly the experimental data points ($\chi^2_{\rm{red}} \approx 1$).

\subsection{Comparison to predictions calculated with global $\alpha$ + nucleus optical potential}
\label{subsec:ang_theo}

In the framework of the $\gamma$ process network studies more than 2 $\times$ 10$^4$ reactions on about 2000 mostly unstable nuclei are taken into account. In many of those reactions alpha particles are present either in the entrance or in the exit channel. Experimentally, most of these reactions are inaccessible for cross section measurements, since short-lived unstable nuclei are involved in the entrance channel. Therefore, global $\alpha$ + nucleus optical potentials are used in the reaction network to predict the needed reaction rates. 

The variation of the potential parameters of the real part as a function of mass and energy is smooth and relatively well understood. On the contrary, the imaginary part of the optical potential is strongly energy-dependent especially at energies around the Coulomb barrier. As a result of these different energy dependencies of the parameterizations, different predictions for reaction cross sections will occur, in particular at very low energies far below the Coulomb barrier. 

In the following, we will compare our new experimental data to the predictions of well-known or recent open-access global potentials.

(I.) In the middle of the 1960s, alpha elastic scattering experiments at E$_{\alpha}$ = 24.7 MeV on nuclei ranging from O to U were performed and a consistent optical potential study was carried out. Based on this study, a global $\alpha$ + nucleus potential was formed. It is a very simple, four-parameter Woods-Saxon potential with mass and energy-independent parameters referred to as the potential of McFadden and Satchler \cite{McFa66}. This potential is the default for the H-F calculations of astrophysical reaction rates in the NON-SMOKER code \cite{Raus00, Raus01}. Despite its simplicity it provides a reasonable description of alpha scattering data and cross sections of $\alpha$-induced reactions, at energies slightly above the Coulomb barrier. The H-F calculations based on this McFadden potential have a tendency to overestimate reaction cross sections at energies below the Coulomb barrier, e.g., \cite{Gyur06, Kiss12}. Recently the depth of this potential was modified by adding a Fermi-type function and this modified potential provided excellent predictions for $\alpha$-induced cross sections on $^{141}$Pr and $^{162,166}$Er \cite{Saue11, Kiss14, Glo14}. 

(II.) The "standard" potential of M. Avrigeanu \textit{et al}.~\cite{Avri10} is an $\alpha$ + nucleus optical potential with Woods-Saxon parameterizations for the real and imaginary parts, with a total of nine independent parameters, three for the real potential, depth $V_0$, radius $R_R$ and diffuseness $a_R$, three to describe the imaginary volume
potential $W_V$, $R_V$, $a_V$ and the remaining three to describe the imaginary surface potential $W_S$, $R_S$, $a_S$. All of these parameters are mass and energy dependent and the $\alpha$ + nucleus optical potential was obtained by fitting alpha particle elastic scattering and reaction cross sections on nuclei in the $45 \leq A \leq 209$ mass range, and for $E < 50$ MeV. Consequently, this potential provides a reasonable description for both the scattering and reaction data. However, the extrapolation of such a many-parameter potential to unstable nuclei with extreme neutron-to-proton ratio $N/Z$ may lead to additional uncertainties in the calculation of astrophysical reaction rates.

In recent years, several $\alpha$-induced cross sections on nuclei with mass number above A $=$ 140 \cite{Kiss11b,Raus12,Saue11,Nett13,Kiss14,Glo14} were measured and, furthermore, new experimental alpha elastic scattering angular distributions on even-odd nuclei became available \cite{Kiss13}, thus improvements of this parametrization became possible. It was shown \cite{avr14} that the accuracy of the cross section predictions can be increased using an updated surface imaginary-potential depth $W_D$. 
However, for the $^{106}$Cd target under study in this work, the modifications in the latest version \cite{avr14} are very minor and restricted to energies below about 10\,MeV, i.e., the modifications do not affect the analysis of the elastic scattering data. Therefore, we show the results using the potential from \cite{Avri10}. 

(III.) Finally, the most recent $\alpha$ + nucleus optical potential used in this work is the ATOMKI-V1 \cite{Mohr13} optical potential parameter set. It is based entirely on alpha elastic scattering angular distributions measured in the  $89 \leq A \leq 144$ mass region at energies above and below the Coulomb barrier. The real part $V(r)$ consists of a double folding parametrization $V_F(r)$, based on the widely used DDM3Y interaction \cite{Mohr97,Satc79,Atzr96} and modified by a strength parameter $\lambda$ derived from the energy-independent volume integral $J_R$. The $W_S$ imaginary part is parametrized by a surface Woods-Saxon function with fixed $R_S$ and $a_S$ geometry and energy-dependent strength $W_S$ and volume integral $J_I$.

The ATOMKI-V1 potential has been constructed for low energies where a pure surface Woods-Saxon potential in the imaginary part is sufficient. Thus, it is not surprising that ATOMKI-V1 is not able to reproduce the angular distributions at the highest energies studied in this work.

Figure \ref{fig:cd106_all_pot_ND} shows the available experimental angular distributions in comparison with the predictions calculated using the above mentioned global optical potential parameter sets. It can be seen that each of the different $\alpha$ + nucleus optical potentials used in this work provides a slightly different description of the experimental data, however, none of them describes the entire experimental data set. For a strict comparison between the potentials, the $\chi^2_{red}$ values and total reaction cross sections $\sigma_{reac}$ can be found in Table \ref{tab:local_pot_parameters}. 

\begin{table}[ht!]
\caption{
\label{tab:table_chi_global}
$\chi^2_{red}$ analysis of the global optical potentials models considered in this work. Please note that the $\star$ indicates that the $\chi^2_{red} >$ 100 and ND refers to the Notre Dame measurements.
}
\begin{adjustbox}{max width=\textwidth}
\begin{tabular}{llllllllll}
\hline
{\bf Global Optical } & \multicolumn{3}{c}{\bf Elastic Scattering ($E_{C.M.}$)} \\
{\bf Potential Model} & {\bf 15.6 MeV - Atomki} & {\bf 16.4 MeV - ND} & {\bf 17.0 MeV - Atomki}  \\
\hline
Avrigeanu {\it et al.} 2010 	& 4.3 & 25.7 & 2.7   \\ 
McFadden \& Satchler 		& 35.1  & $\star$ 
				& 41.9  \\
ATOMKI-V1 			& 7.7 & 1.8 & 5.9  \\ 
Avg. local potential 		& 4.1 & 1.7 & 3.9 \\
\hline 
{} & {\bf 17.0 MeV - ND} & {\bf 18.3 MeV - ND} & {\bf 18.9 MeV - Atomki}  \\
\hline
Avrigeanu {\it et al.} 2010 	& 27.8 & 22.1 & 5.2 \\ 
McFadden \& Satchler 		& $\star$ 
				& $\star$  
				& 72.0 \\ 
ATOMKI-V1 			& 1.6 & 1.7 & 1.3 \\
Avg. local potential 		& 2.2 & 7.3 & 4.9 \\
\hline
{ } & {\bf 21.2 MeV - ND$\dag$} & {\bf 23.6 MeV - ND$\dag$} & {\bf 26.0 MeV - ND$\dag$} & {} \\
\hline
Avrigeanu {\it et al.} 2010 	& 19.7 & 20.7 & $\star$ 
\\
McFadden \& Satchler 		& $\star$  
				& $\star$ 
				& $\star$ 
				\\ 
ATOMKI-V1 			& 27.2 & $\star$ 
				& $\star$ 
				\\
Avg local potential & $\star$  
& $\star$  
& $\star$ 
\\
\hline
$\dag$ - Abs. renorm. & 1.016  & 0.959  & 0.921 \\ 
\hline
\end{tabular}
\end{adjustbox}
\end{table}

From the results of Table \ref{tab:table_chi_global}, on the one hand, we found that the considered global $\alpha$ + nucleus optical potentials roughly describe the experimental data in the energy range from E$_{C.M.}$ = 15.6 MeV to E$_{C.M.}$ = 18.9 MeV. On the other hand, above E$_{C.M.}$ = 18.9 MeV, all $\alpha$ + nucleus optical potentials fail to accurately describe the data, providing relatively large $\chi^2_{\rm{red}}$ values. This asks for further improvement of global \anucpots\ but does not affect the following study of low-energy \al -induced reaction cross sections.

\section{Comparison to $\alpha$-induced reaction cross sections on $^{106}$Cd} 
\label{sec:capture_reac}
Reaction cross sections of \al -induced reactions for heavy nuclei can be calculated using the statistical model (StM) \cite{Haus52}. In particular, this model has been widely applied for the calculation of reaction cross sections and stellar reaction rates in nuclear astrophysics \cite{Rau11}. In a schematic notation, the reaction cross section in the StM is proportional to
\begin{equation}
\sigma(\alpha,X) \sim \frac{T_{\alpha,0} T_X}{\sum_i T_i}
\label{eq:StM}
\end{equation}
with the transmission coefficients $T_i$ into the $i$-th open channel. The $T_i$ are calculated from global optical potentials (particle channels) and from the $\gamma$-ray strength function for the photon channel. For details of the definition of $T_i$, see \cite{Rau11}. $T_{\alpha,0}$ refers to the entrance channel where the target nucleus is in its ground state. Note that this holds only for laboratory experiments; under stellar conditions, i.e.,\ for the calculation of stellar reaction rates $N_A < \sigma v >$, thermal excitations of the target have to be considered. 

It is typical for \al -induced reactions on heavy nuclei that $T_{\alpha}$ is much smaller than the other $T_i$. A simple qualitative explanation is the high Coulomb barrier in the \al\ channel. In the neutron channel, a Coulomb barrier is completely missing and, in the proton channel, the barrier is much lower. As a consequence, the cross section in the StM in Eq.~(\ref{eq:StM}) factorizes into a production cross section of the compound nucleus which is proportional to $T_{\alpha,0}$, and a decay branching ratio $b_X = T_X/\sum_i T_i$ practically independent of $T_\alpha$. The production cross section is thus entirely defined by the chosen $\alpha$ + nucleus optical potential whereas the branching ratio $b_X$ depends practically not on the $\alpha$ + nucleus optical potential but on all the other ingredients of the StM (optical potentials for the other channels, $\gamma$-ray strength functions, level densities). Consequently, all cross sections of \al -induced reactions are sensitive to the $\alpha$ + nucleus optical potential, but each individual \rap , \ran , or \rag\ reaction has further and complicated sensitivities to the other ingredients. As the $\alpha$ + nucleus optical potential affects directly the production cross section, the sensitivity (as defined e.g.\ in \cite{Raus12a}) is close to 1 for all (\al ,$X$) reactions at energies around or below the Coulomb barrier.

The reaction $Q$-values for \al -induced reactions on \Cd\ are $-10.15$\,MeV for the \ran , $-5.51$\,MeV for the \rap , and $+1.13$\,MeV for the \rag\ reactions. Experimental data for these reactions are available at low energies from about 7\,MeV to 13\,MeV \cite{Gyur06}. Following the above argumentation, it can be expected that the \ran\ cross section is dominating at higher energies down to close above the \ran\ threshold. Below 10\,MeV, only the \rap\ and \rag\ channels are open, and it turns out that the \rap\ channel becomes very weak towards lower energies.

\begin{figure}[ht!]
\begin{center}
\includegraphics[ width =85mm , clip]{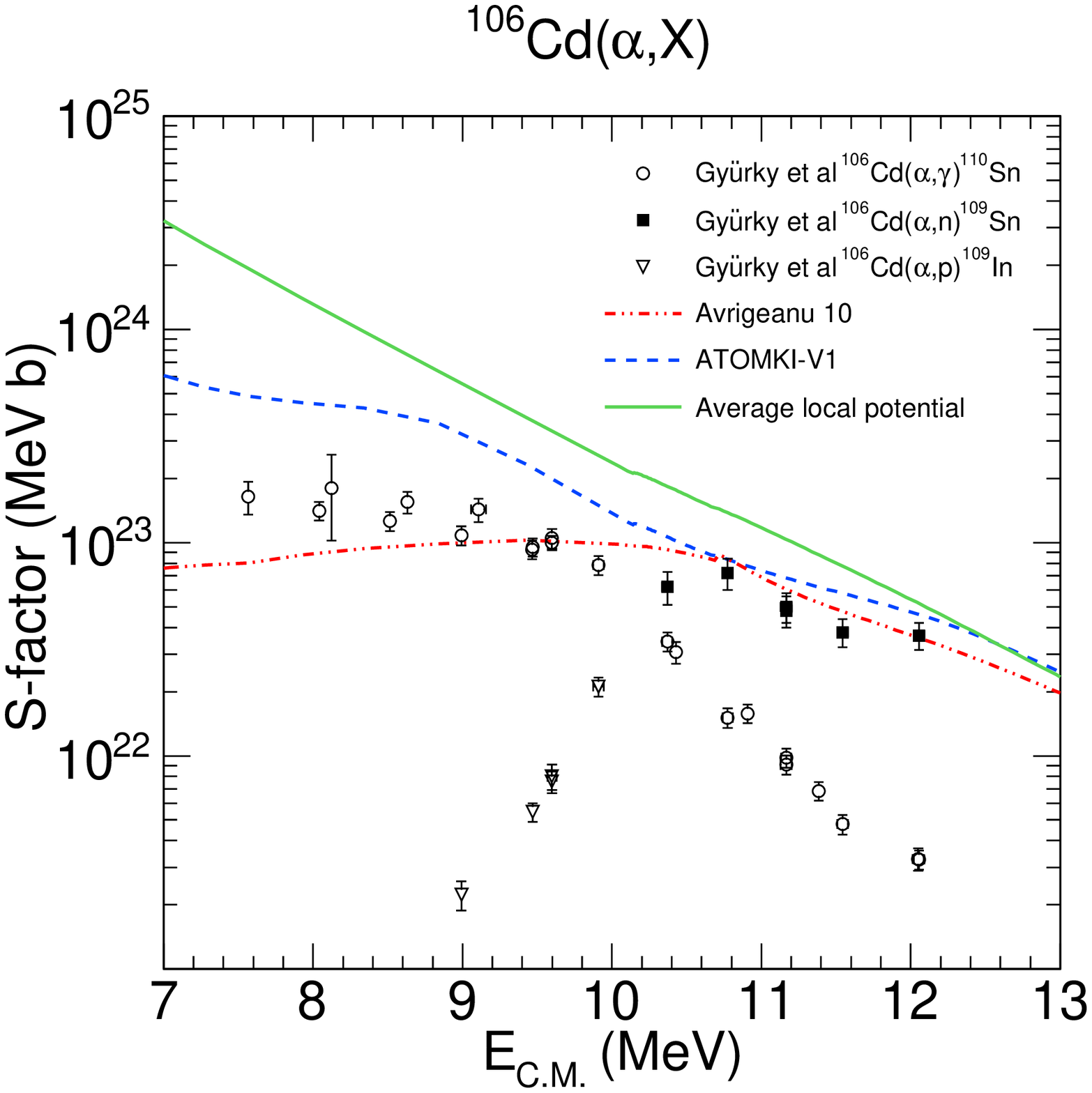}
\caption{
\label{fig:total}
(Colour online) Total reaction cross
section for \al\ $+$ \Cd\ from different potentials, converted to the astrophysical
S-factor, and compared compared to the experimental \rag , \rap , and \ran\ cross
sections \cite{Gyur06}. Further discussion, see text.
}
\end{center}
\end{figure}

The focus of the present study is the $\alpha$ + nucleus optical potential for \Cd\ which determines the compound formation cross section. In Fig.~\ref{fig:total}\footnotemark, we show the formation cross section taken as the sum over the dominating \rag , \ran , and \rap\ channels for the potentials under study in this work. Obviously, the calculated compound formation cross section should be an upper envelope of the experimental \rag , \ran , and \rap\ data. At higher energies above about 13\,MeV, all potentials lead to very similar cross sections. This is a general finding which has been explained from the properties of reflexion coefficients $\eta_L$, see \cite{Mohr11,Mohr13a}. However, at lower energies, significant differences can be seen. The local potential which has been adjusted to the elastic scattering angular distibutions overestimates the experimental data at energies below about 10\,MeV. Unfortunately, it was not possible to determine the energy dependence of the imaginary part from elastic scattering and, therefore, the local potential behaves similar to the widely used potential by McFadden and Satchler \cite{McFa66} which has also been determined from elastic scattering at higher energies and also has an energy-independent imaginary part. The reaction cross sections from the McFadden/Satcher potential have already been shown in \cite{Gyur06}.

\footnotetext{In astrophysical investigations it is common to quote the astrophysical $S$ factor \cite{ilibook}. The cross section, $\sigma$(E) and the astrophysical $S$ factor, $S$(E) at c.m. energy $E$ are related by:

\begin{equation}
 S(E)=E\sigma(E) exp(2\pi\eta), 
\end{equation}

with $\eta$ being the Sommerfeld parameter. For better visibility, in the figures instead of the cross sections, the astrophysical S factors will be used throughout the present manuscript.}

The new ATOMKI-V1 potential uses a very similar parametrization as the local potential, but it has an energy-dependent imaginary part in addition. The decreasing imaginary strength towards lower energies leads to smaller cross sections, but the calculation still overestimates the experimental data. On the contrary, the potential by Avrigeanu {\it et al.}\ \cite{Avri10} underestimates the data. The latest revision of this potential \cite{avr14} has a slightly stronger imaginary part at very low energies which leads to slightly larger cross sections by about 10\,\% below 10\,MeV.

Let us now turn to the individual \rag , \ran , and \rap\ reactions. As already stated above, these cross sections depend not only on the $\alpha$ + nucleus optical potential with sensitivity ${\hat{S}} \approx 1$ as defined in \cite{Rau11}, but also on the further ingredients of the StM with varying energy-dependent sensitivities ${\hat{S}}$. This complicates the comparison of the results of the StM with the experimental data. The following calculations are the results from SMARAGD code \cite{Rau09} with its default parameters. Following the scope of this study, only the $\alpha$ + nucleus optical potential has been varied.

\begin{figure}[ht!]
\begin{center}
\includegraphics[ width = 80 mm, clip]{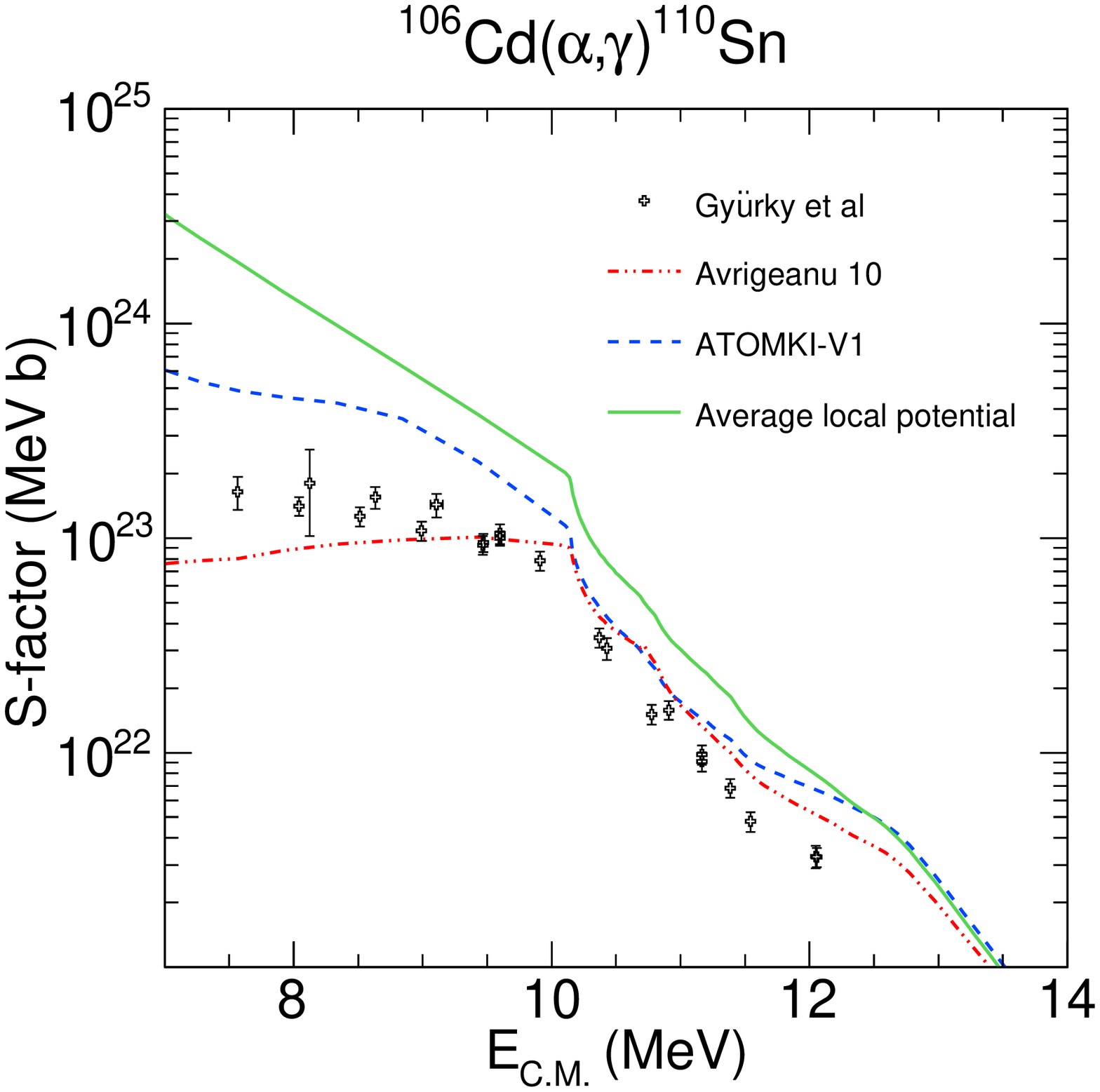}
\caption{
\label{fig:106cd_ag_others}
(Colour online) Astrophysical S-factor for the \Cd \rag \Sn\ reaction, experimental data from \cite{Gyur06}. 
The figure contains the predictions from global $\alpha$ + nucleus potentials~\cite{Mohr13,Avri10} and the average local potential.
 } 
\end{center}
\end{figure}

Fig.~\ref{fig:106cd_ag_others} shows the astrophysical S factors of the \Cd \rag \Sn\ reaction which dominates below 10\,MeV. Here the \rag\ cross section is essentially defined by the $\alpha$ + nucleus optical potential. Similar to the compound formation cross section in Fig.~\ref{fig:total}, the local potential overestimates the low-energy data significantly, the ATOMKI-V1 potential slightly overestimates, and the potential by Avrigeanu {\it et al.}\ slightly underestimates. At higher energies above the neutron threshold, the cross sections become $\sigma(\alpha,\gamma) \sim T_{\alpha,0} T_\gamma / \sum_i T_i$ with $\sum_i T_i \approx T_n$ and, thus, no direct conclusion on the $\alpha$ + nucleus optical potential is possible here. As a word of caution, it must be mentioned that the behavior of the \rag\ cross section at low energies may also be affected by direct reactions, in particular, Coulomb excitation of the first $2^+$ state in \Cd\ \cite{Rau13}.

\begin{figure}
\begin{center}
\includegraphics[ width = 80 mm, clip]{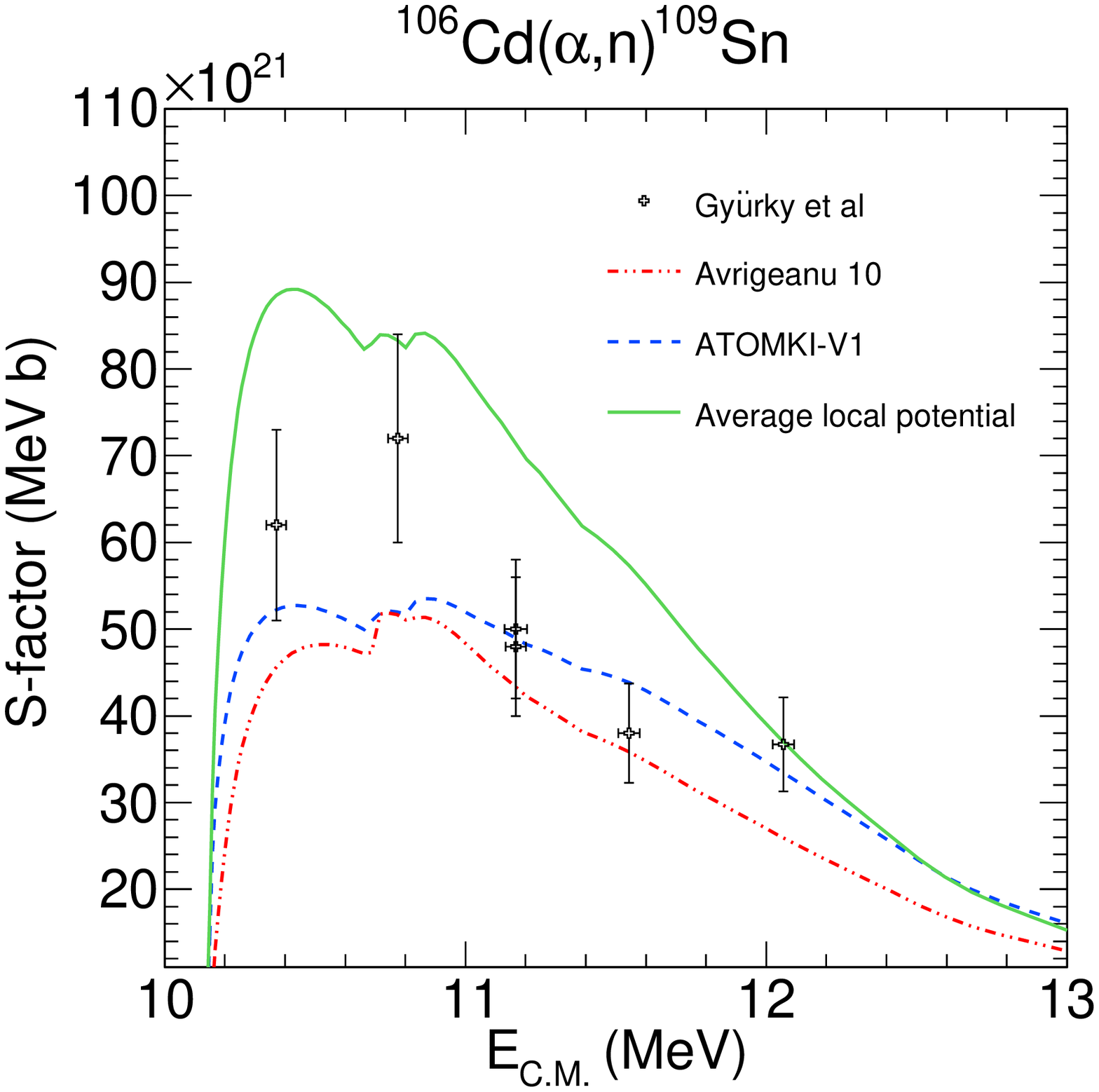}
\caption{
\label{fig:106cd_an_others}
(Colour online) Astrophysical S-factor for the \Cd \ran \Snn \ reaction, experimental data from \cite{Gyur06}.  
The figure contains the predictions from global $\alpha$ + nucleus potentials~\cite{Mohr13,Avri10} and the average local potential.
}
\end{center}
\end{figure}

Fig.~\ref{fig:106cd_an_others} shows the same comparison for the \Cd \ran \Snn\ reaction. At energies sufficiently above the threshold, cross sections become the $\sigma(\alpha,n) \sim T_{\alpha,0} T_n / \sum_i T_i \approx T_{\alpha,0}$ for $\sum_i T_i \approx T_n$. In principle, it would be possible to derive properties of the $\alpha$ + nucleus optical potential from these data; however, the experimental data cover only a narrow energy interval of about 2\,MeV from 10\,MeV to 12\,MeV, and the error bars are not sufficiently small to allow clear conclusions.

\begin{figure}[ht!]
\begin{center}
\includegraphics[ width = 80 mm, clip]{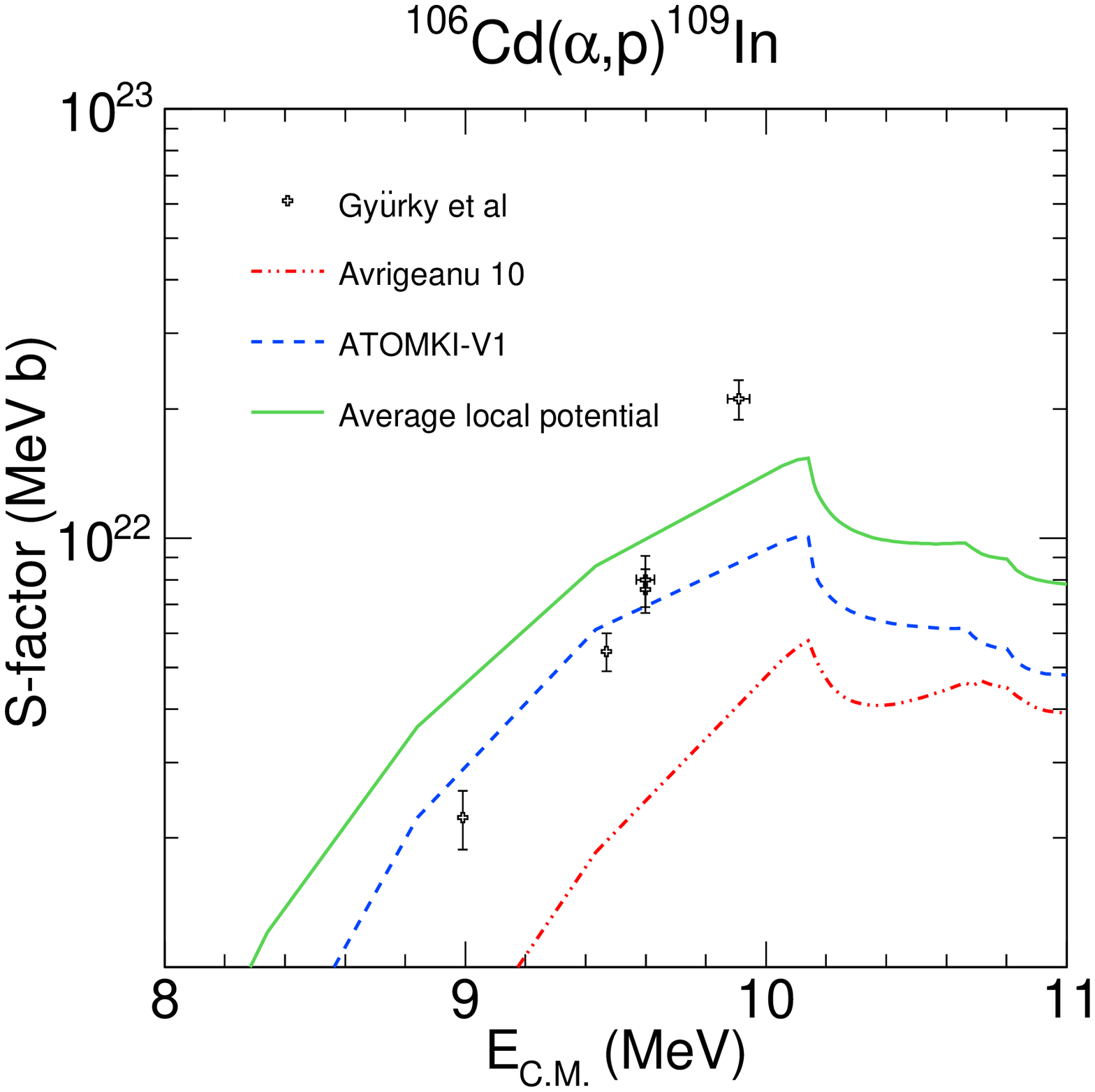}
\caption{
\label{fig:106cd_ap_others}
(Colour online) Astrophysical S-factor for the \Cd \rap \In \ reaction, experimental data from \cite{Gyur06}. 
The figure contains the predictions from global $\alpha$ + nucleus potentials~\cite{Mohr13,Avri10} and the average local potential.
}
\end{center}
\end{figure}

Fig.~\ref{fig:106cd_ap_others} shows the astrophysical S factors of the \Cd \rap \In\ reaction. Similar to the \ran\ data, the energy range is also very limited (9\,MeV - 10\,MeV). As the cross section becomes $\sigma(\alpha,p) \sim T_{\alpha,0} T_p / \sum_i T_i \approx T_{\alpha,0} T_p/(T_p + T_\gamma)$ in the energy range of the experimental data, it is not possible to directly conclude on the $\alpha$ + nucleus optical potential from these data.

For completeness it should be mentioned here that the study of cross section ratios like, e.g.\ $\sigma(\alpha,p)/\sigma(\alpha,\gamma) \sim \big(T_{\alpha,0} T_p / \sum_i T_i \big) / \big( T_{\alpha,0} T_\gamma / \sum_i T_i \big) = T_p/T_\gamma$ allows to constrain the other ingredients of the StM except the $\alpha$ + nucleus optical potential. For an application of this idea, see e.g.\ \cite{Kiss13}.
As we have seen (and has been stated in our earlier study of the reaction data \cite{Gyur06}), it is not straightforward to draw conclusions on the $\alpha$ + nucleus optical potential from the cross sections of the \rag , \ran , and \rap\ reactions.

\section{Summary and conclusions}
\label{sec:summary}
We have measured three angular distributions of \Cd \raa \Cd\ elastic scattering at energies around the Coulomb barrier; and we have analyzed these data and data from the literature \cite{Palu12} within the optical model. Reasonable descriptions can be obtained for all data using a real folding potential with two adjustable parameters (strength $\lambda$ and width $w$) and an imaginary part of Woods-Saxon type. Similar to previous results \cite{Gala05}, at low energies, a surface Woods-Saxon type parametrization is sufficient, whereas at higher energies a combination of volume and surface imaginary part improves the obtained $\chi^2_{red}$ significantly. A local potential is derived from the average parameters. Unfortunately, it was not possible to determine the energy dependence of the imaginary part at low energies. The new experimental data are also described very reasonably by the recent global potentials \cite{Mohr13,Avri10,avr14}, whereas the old McFadden/Satchler potential \cite{McFa66} has a trend to overestimate the elastic data.

The locally adjusted potential and the global potentials by Avrigeanu {\it et al.}\ \cite{Avri10} and ATOMKI-V1 \cite{Mohr13} were used to calculate the cross sections of \al -induced reactions on \Cd . It is shown that these cross sections factorize into a compound production cross section which depends only on the chosen $\alpha$ + nucleus optical potential and into a decay branching which is practically independent of the $\alpha$ + nucleus optical potential but shows a complicated dependence on the other ingredients of the StM. Therefore, clear conclusions on the $\alpha$ + nucleus optical potential from \al -induced reaction data are difficult in the case of \Cd . Nevertheless, some conclusions can be drawn. It turns out that the locally adjusted potential overestimates the reaction data at low energies. This result is similar to the one obtained with the widely used McFadden/Satchler potential and can be attributed to the missing energy dependence of the imaginary part. The predictions from the new global potentials \cite{Mohr13,Avri10,avr14} are much closer to the experimental data \cite{Gyur06} but still far from an excellent description with $\chi^2_{\rm{red}} \approx 1$. The ATOMKI-V1 potential \cite{Mohr13} slightly overestimates the reaction data at low energies whereas the potential by Avrigeanu {\it et al.}\ underestimates the data. As similar results have been found for $^{140}$Ce and $^{141}$Pr, the best prediction of \al -induced reaction cross sections in the mass range $100 \le A \le 150$ could be taken from the geometric mean of these recent global potentials.

\section*{Acknowledgments}

This work was supported by OTKA (K101328, K108459), by the Hungarian \& Portuguese Intergovernmental S\&T Cooperation Programme NO. T\'ET\_10-1-2011-0458 and by the ENSAR/THEXO European FP7 programme. 
G. G. Kiss acknowledges support from the J\'anos Bolyai Research Scholarship of the Hungarian Academy of Sciences. T. Rauscher acknowledges support from the Swiss NSF, the UK STFC grant BRIDGCE (ST/M000958/1) and the European Research Council. K. Sonnabend acknowledges support from DFG (SO907/2-1).


\begin{thebibliography}{}
%
\bibitem{kap11} F. K\"appeler {\it et al.,} Rev. Mod. Phys. 83 (2011) 157.
%
\bibitem{rei14} R. Reifarth {\it et al.,} J. Phys. G 41 (2014) 053101.
%
\bibitem{Arno14}
M. Arnould, S. Goriely, The r-Process of Nucleosynthesis: The Puzzle is Still with Us: Astrophysics,
Ed. Prof. Ibrahim Kucuk, ISBN: 978-953-51-0473-5, InTech (2014),
available from: http://cdn.intechopen.com/pdfs-wm/34260.pdf
%
\bibitem{Arg04} D. Argast {\it et al.,} Astronomy and Astrophysics 416 (2004) 997
%
\bibitem{Woos78} 
S. E. Woosley and W. M. Howard, Astrophys. J. Suppl 36 (1978) 285.
%
\bibitem{Arno03} 
M. Arnould and S. Goriely, Phys. Rep. 384 (2003) 1.
%
\bibitem{Kus11} M. Kusakabe {\it et al.,} Astrophys. J. 726 (2011) 25.
%
\bibitem{arl99} 
C. Arlandini {\it et al.,}  Astrophys. J. 525 (1999) 886.
%
\bibitem{nem94}
Zs. N\'emeth {\it et al.,} Astrophys. J. 426 (1994) 357.
%
\bibitem{Rau13a}
T. Rauscher {\it et al.,} Rep. Prog. Phys. 76 (2013) 066201
%
\bibitem{Trav11}
C. Travaglio {\it et al.,}  The Astrophysical Journal 739 (2011) 93.
%
\bibitem{Trav14}
C. Travaglio {\it et al.,} The Astrophysical Journal 795 (2014) 141.
%
\bibitem{Scha98}
H. Schatz {\it et al.,} Physics Reports 294 (1998) 167.
%
\bibitem{Froh06a}
C. Fr\"ohlich {\it et al.,}  Phys. Rev. Lett. 96 (2006) 142502.
%
\bibitem{Howa91}
W. M. Howard, B. S. Meyer and S. E. Woosley, The Astrophysical Journal Letters 373 (1991) L5.
%
\bibitem{Raus02}
T. Rauscher, A. Heger, R. D. Hoffman and S. E. Woosley, The Astrophysical Journal 576 (2002) 323.
%
\bibitem{Raye95}
M. Rayet {\it et al.,} Astronomy and Astrophysics 298 (1995) 517.
%
\bibitem{Raus97}
T. Rauscher, F. K. Thielemann and K. L. Kratz, Phys. Rev. C 56 (1997) 1613.
%
\bibitem{Raus00}
T. Rauscher and F. K. Thielemann, At. Data Nucl. Data Tables 75 (2000) 1.
%
\bibitem{Kiss11b}
G. G. Kiss {\it et al.,} Physics Letters B 695 (2011) 419.
%
\bibitem{Raus12}
T. Rauscher {\it et al.,} Phys. Rev. C 86 (2012) 015804.
%
\bibitem{Saue11}
A. Sauerwein {\it et al.,} Phys. Rev. C 84 (2011) 045808. 
%
\bibitem{Nett13}
L. Netterdon {\it et al.,} Nuclear Physics A 916 (2013) 149.
%
\bibitem{Kiss14}
G. G. Kiss {\it et al.,} Physics Letters B 735 (2014) 40.
%
\bibitem{Glo14} J. Glorius {\it et al.,} Phys. Rev. C 89 (2014) 065808.
%
\bibitem{Palu12}
A. Palumbo {\it et al.,} Phys. Rev. C 85 (2012) 035808. 
%
\bibitem{Gyur06} 
Gy. Gy\"urky {\it et al.,} Phys. Rev. C 74 (2006) 025805.
%
\bibitem{Mohr13}
P. Mohr {\it et al.,} At. Data Nucl. Data Tables 99 (2013) 651.
%
\bibitem{Mohr97}
P. Mohr {\it et al.,} Phys. Rev. C 55 (1997) 1523.
%
\bibitem{Fulo01}
Zs. F\"ul\"op {\it et al.,} Phys. Rev. C 64 (2011) 065805.
%
\bibitem{Gala05}
D. Galaviz {\it et al.,} Phys. Rev. C 71 (2005) 065802.
%
\bibitem{Kiss06}
G. G. Kiss {\it et al.,} European Physical Journal 27 (2006) 197.
%
\bibitem{Kiss08}
G. G. Kiss {\it et al.,} Journal of Physics G 35 (2008) 014037.
%
\bibitem{Kiss09}
G. G. Kiss {\it et al.,} Phys. Rev. C 80 (2009) 045807.
%
\bibitem{Kiss11}
G. G. Kiss {\it et al.,} Phys. Rev. C 83 (2011) 065807.
%
\bibitem{Kiss13}
G. G. Kiss {\it et al.,} Phys. Rev. C 88 (2013) 045804.
%
\bibitem{Gala04} 
D. Galaviz, PhD thesis, 2006 Technische Universit\"at Darmstadt
%
\bibitem{Abel93}
H. Abele and G. Staudt, Phys. Rev. C 47 (1993) 742.
%
\bibitem{Vrie87}
H. de Vries, C. W. de Jager and C. de Vries, At. Data Nucl. Data Tables 36 (1987) 495.
\bibitem{Satc79}
R. G. Satchler and G. W. Love, Physics Reports 55 (1979) 183. 
%
\bibitem{Atzr96}
U. Atzrott {\it et al.,} Phys. Rev. C 53 (1996) 1336.
%
\bibitem{Avri09}
M. Avrigeanu {\it et al.,} At. Data Nucl. Data Tables 95 (2009) 501.
%
\bibitem{DFOLD}
H. Abele,
University of T\"ubingen, 
{computer code DFOLD} (unpublished).
%
\bibitem{Chis96}
V. Chist\'e {\it et al.,} Phys. Rev. C 54 (1996) 784.
%
\bibitem{McFa66} 
L. McFadden and G. R. Satchler, Nucl. Phys. 84 (1966) 177. 
%
\bibitem{Raus01} 
T. Rauscher and F. K. Thielemann, At. Data Nucl. Data Tables 79 (2001) 47.
%
\bibitem{Kiss12} 
G. G. Kiss {\it et al.,} Phys. Rev. C 86 (2012) 035801.
%
\bibitem{Avri10}
M. Avrigeanu and V. Avrigeanu, Phys. Rev. C 82 (2010) 014606. 
%
\bibitem{avr14}
V. Avrigeanu {\it et al.,} Phys. Rev. C 90 (2014) 044612.
%
\bibitem{ilibook} C. Iliadis, {\it Nuclear Physics of Stars} (Wiley, Weinheim 2007).
%
\bibitem{Haus52}
W. Hauser and H. Feshbach, Phys. Rev. 87 (1952) 366.
%
\bibitem{Rau11}
T. Rauscher, Int. J. Mod. Phys. E 20 (2011) 1071.
%
\bibitem{Raus12a}
T. Rauscher, The Astrophysical Journal Supplement Series 201 (2012) 26.
%
\bibitem{Mohr11}
P. Mohr, Phys. Rev. C 84, (2011) 055803.
%
\bibitem{Mohr13a}
P. Mohr, Phys. Rev. C 87 (2013) 035802.
%
\bibitem{Rau09}
T. Rauscher
Computer code SMARAGD, (2009) http://nucastro.org/smaragd.html
%
\bibitem{Rau13}
T. Rauscher, Phys. Rev. Lett. 111 (2013) 061104.
%
\end{thebibliography}
\end{document}